\documentstyle[preprint,eqsecnum,aps]{revtex}
\begin{document}
\draft
\title{Classical and quantum decay of one dimensional\\
finite wells with oscillating walls.}
\author{A.J.Fendrik\cite{one} and D.A.Wisniacki\cite{two}}
\address{Departamento de F\'{\i}sica, Facultad de Ciencias Exactas y 
Naturales,\\
Universidad de Buenos Aires, Ciudad Universitaria, 1428, Buenos Aires, \\
Argentina.}
\date{\today}
\maketitle

\begin{abstract}
To study the time decay laws (tdl) of quasibounded hamiltonian systems we
have considered two finite potential wells with oscillating walls filled by
non interacting particles. We show that the tdl can be qualitatively
different for different movement of the oscillating wall at classical level
according to the characteristic of trapped periodic orbits. However, the
quantum dynamics do not show such differences.
\end{abstract}


\narrowtext

\newpage

\section{Introduction}

\label{sec:int}

The time decay laws (tdl) of classical quasibounded hamiltonian systems use
to include meaningful information related to the bounded transient dynamics.
The feature of the tdl corresponding to a fully chaotic system has been well
established. In a special way, the tdl's are greatly sensible to the
characteristic of the trapped periodic orbits. Systems having a fully
hyperbolic set of trapped periodic orbits show a purely exponential tdl,
while the tdl becomes algebraic if they have a parabolic subset of them. So,
in the case of the Sinai billiard\cite{Sinai} and the Bunimovich stadium 
\cite{Buni}, paradigms of fully chaotic hamiltonian systems, the existence
of the so called bouncing ball orbits (parabolic and non isolated periodic
orbits) gives algebraic long time tails in the tdl's
\cite{Majo1,Majo2,German}.

On the other hand, it is a well known fact that the characteristic and the
distribution of periodic orbits in bounded systems are relevant subjects to
quantify semiclassically such systems\cite{Gutz}.

To establish the influence of the periodic orbits in the decay of quantum
systems, in the present work we have studied the decay processes in two
quasibounded systems whose trapped periodic orbits have different
characteristics.

We show that the classical systems having trapped periodic orbits of
different characteristics have qualitatively different tdl, while the
quantum analogues do not show such sensitivity. In other words, we show that
two classical systems with exponential and algebraic tdl respectively they
have algebraic quantum tdl (qtdl) when they are considered as quantum
systems.

Our work is organized in the following way. In Sec.\ \ref{sec:pres} we
introduce the classical systems whose tdl we study in Sec.\ \ref{sec:tdl}.
Sec.\ \ref{sec:num} is devoted to describe the quantum systems and the
numerical method that we have employed to compute the qtdl. In Sec.\ \ref
{sec:decay} we show the resulting qtdl. Finally, Sec.\ \ref{sec:con} is
devoted to conclusions. We include one appendix to show some characteristics
of the tdl's.

\section{THE CLASSICAL SYSTEMS}

\label{sec:pres}

In the present work we deal with a kind of one-dimensional periodically
driven hamiltonians. The corresponding system can be described in the
extended phase space\cite{Lich} as a conservative system with two degrees of
freedom.

Let us consider non interacting point particles of unity mass moving inside
a one-dimensional time dependent potential well

\begin{eqnarray}
V(q,t) = \left\{ 
\begin{array}{ll}
\infty & \mbox{if $q < 0$} \;, \\ 
0 & \mbox{if $0 \leq q \leq a(t)$ } \;, \\ 
\infty & \mbox{if $ a(t) < q$ } \;. \label{pozo}
\end{array}
\right.
\end{eqnarray}

That is a perfectly reflecting one-dimensional box with an oscillating wall
according to the time dependent law $a(t)$.

Looking at the graph of the position of one particle $q$, as a function of
time $t$, we can see that the system is equivalent to a particle forwards in
a two dimensional infinite pipe with appropriate reflecting conditions at
the boundary (See Fig. ~\ref{cano}(a)) . The reflecting condition (the
change of sign of the relative velocity between the walls and the particle)
imposes:

\begin{eqnarray}
\begin{array}{ll}
v_{f}= -v_{i} & \mbox{at $q=0$} \;, \\ 
v_{f}=2 \dot{a}(t) - v_{i} & \mbox{ at $q=a(t)$} \;, \label{reflex}
\end{array}
\end{eqnarray}

where $v_{f} (v_{i})$ is the velocity of the particle after (before) the
collision. Both velocities correspond to the sloops of the straight lines in
the $q.vs.t$ graphs ($v=\tan{\alpha}$).

Let be $a(t)$ periodic in time, so $a(t)=a(t+\tau)$.

Taking account the time lattice translation symmetry we can consider only an
elemental segment of the channel setting periodic boundary conditions,
namely $t=\tau \rightarrow t=0$ so that $q(t=\tau) \rightarrow q(t=0)$ and $
v(t=\tau) \rightarrow v(t=0)$. So, the system can be seen as a two
dimensional billiard where particles leaving at $t=\tau$ emerge at $t=0$
having the same velocity $v$ and the same position $q$ while the collisions
at the walls follows the laws (\ref{reflex}). In this version of the system,
we have two degrees of freedom associated to the coordinates $q$ and $
\theta= t-[t/\tau] \tau$ (reduced time) where $[...]$ means the integer part
(See Fig. ~\ref{cano} (b)). The conjugate momenta are $v$ and $-E$ (kinetic
energy) respectively.

When the potential well is given by rule (\ref{pozo}) particles inside the
box are bounded despite their velocities. On the other hand, if the
oscillating wall (at $q=a(t)$) is a finite barrier of potential of height 
$V_{0}$, particles with velocities $\mid v \mid > \sqrt{2V_{0}}$ will leave
the well when they reach the wall. In this case, according to the
characteristic of the dynamics, which is controlled by the law $a(t)$ of the
oscillating wall, the system can be transiently bounded.

We have focused our attention in two particular rules $a(t)$. In the
following we take $\tau$ as a unit of time and the mean value of $a(t),
<a>_{t}=1/\tau \int_{0}^{\tau} a(t) dt$ as a unit of length and we take the
energies as dimensionless (that is they are divided by $m <a>^{2}_{t} /
\tau^{2}$).

We have considered:

a) Harmonic oscillations

That is to say:

\begin{equation}
a(t)= 1+ \delta \sin{2 \pi t}
\end{equation}

This system resembles the so called full Fermi accelerator model. Such
system was widely studied in simplified versions\cite{Lich}. Figure ~\ref
{fermsec} shows the Poincare surface section for $q=0$ corresponding to the
kinetic energy $E$ versus the $\theta$. Here we can distinguish three
regions. The low energy region is fully chaotic, the intermediate energy
region shows mixed dynamics (regular island surrounded by the chaotic sea)
and finally the adiabatic region, where the dynamics become regular and we
can see invariant curves.

Now we assume that the moving wall is a finite barrier of potential of
height $V_{0}$. So, the particles can leave the well and the limit between
the bounded motion and the unbounded one is the line $E=V_{0}$.

If $V_{0}$ lies on the adiabatic region, that is there is an invariant curve
below the limit line, the system will remain bounded. So to study the tdl we
will set $V_{0}$ less than the lower invariant curve.

b) Sawtooth oscillations

In this case: 
\begin{eqnarray}
a(t) = \left\{ 
\begin{array}{ll}
(1 + 4 \delta t) & \mbox{if $ 0 \leq t \leq 1/4 $ } \;, \\ 
(1+ 2 \delta) - 4 \delta t & \mbox{if $ 1/4 < t \leq 3/4 $ } \;, \\ 
(1- 4 \delta) + 4 \delta t & \mbox{if $ 3/4 < t < 1 $ } \;. \label{saw}
\end{array}
\right.
\end{eqnarray}

The main feature of this system is that the periodic orbits are parabolic
and non isolated. The two dimensional equivalent billiard is like billiards
with neutral boundaries as polygonal ones\cite{Romb,Vega}. For example, Fig.
~\ref{perio} shows a representative periodic orbit. Looking at the billiard
as map ${\cal T}$ 
\begin{eqnarray}
\begin{array}{ccc}
(q(\theta=0),\tan{\alpha(\theta=0)}) & \stackrel{\cal T}{\longrightarrow} & 
(q^{\prime}(1),\tan{\alpha^{\prime}(t=1)})
\end{array}
\end{eqnarray}
the periodic orbit corresponds to a fixed point of $ {\cal T} ^{2}$. A
straightforward calculation shows that ${\cal T} ^{2}$ has eigenvalues 
$\lambda_{1}=\lambda_{2}=1$ when it is linearized at this fixed point. So, it
corresponds to a parabolic periodic orbit. On the other hand, given a fixed
point $(q_{0},\tan{\alpha_{0}})$, it can be shown that there exists an
interval of $q$, $I= (q_{min},q_{max})$, such that $(q,\tan{\alpha_{0}})$ is
a fixed point if $q \epsilon I$. That is, the fixed points are non isolated.

\section{THE DECAY OF CLASSICAL SYSTEMS}

\label{sec:tdl}

To study the tdl, we fill a portion of the bounded region with $N_{0}= 10
^{5}$ particles whose initial conditions are uniformly (random) distributed
in $q$ and $v=\tan{\alpha}$. We compute the ratio $N(t)/N_{0}$ of the
remaining population as function of time.

Let us remark that the initial population must be chosen with low energy (in
the chaotic region) to ensure the decay. If not so, particles can have
initial conditions such that they will remain trapped because, in the case
of harmonic oscillations, they can be on regular island included in the
bounded region. On the other hand, we want to exclude the population of the
region corresponding to very low energy to minimize the effect of the
asymptotic orbits to the (trivial) parabolic subset of periodic orbits
characterized by $E=0$. These trajectories are parallel lines to the time
axis in the two dimensional pipe version of the system and they correspond
to particles that having $v=0$, they will never hit the boundary of the well
(see appendix). So we have considered 
\begin{equation}
E_{min} \leq E \leq E_{max}.
\end{equation}
Figure ~\ref{fermsec2} shows a part of the Poincare surface section of 
Fig. ~\ref{fermsec} and the region that was initially populated. We have 
taken $E_{min}=2.64$ and $E_{max}=5.35$.

Figures ~\ref{cltdl} (a) and (b) show the results corresponding to the
harmonic and sawtooth oscillations in log-lin and log-log plot respectively.
We can see that the curves are quite different. In the case of Fig.~\ref
{cltdl}(a) the log-lin plot shows an almost exponential behavior for the
harmonic oscillations while Fig.~\ref{cltdl}(b) the log-log plot shows an
algebraic one for the sawtooth oscillations. Such a difference can be
explain in terms of the features of the periodic orbits that are included in
the trapped region\cite{Majo1}. In the case of harmonic oscillations, the
trapped periodic orbits are hyperbolic and isolated (aside the subset $E=0$
that was not initially populated) while they are parabolic and non-isolated
(like the bouncing ball orbits of the Sinai billiard) for the sawtooth
oscillations. We stress that the deviation of the exponential behavior in
the case of harmonic oscillations can be attributed to the population of
very low energy region by indirect way (after collisions, see appendix).

\section{QUANTUM SYSTEMS}

\label{sec:num}

To study the quantum analogues of the systems, we want to solve the time
dependent Schroedinger equation. 
\begin{equation}
i \hbar \frac{\partial |\alpha,t_{0},t>}{\partial t} = \hat{H}(q,t)
|\alpha,t_{0},t>
\end{equation}
being $|\alpha,t_{0},t_{0}>$ an initial condition and $\hat{H}(q,t)$ the
Hamiltonian operator $\hat{H}=\hat{T}(\hat{p})+\hat{V}(\hat{q},t)$ where:

\begin{eqnarray}
V(q,t) = \left\{ 
\begin{array}{ll}
0 & \mbox{if $0 \leq q \leq a(t)$ } \;, \\ 
V_{0} & \mbox{if $ a(t) < q$ } \;. \label{pozo1}
\end{array}
\right.
\end{eqnarray}

We have employed an expectral method (fast Fourier transform, FFT) based on
the split of the evolution operator $\hat {U}(t,t_{0})$\cite{Fait1} defined
by 
\begin{equation}
|\alpha,t> = \hat {U}(t,t_{0}) |\alpha,t_{0}>
\end{equation}

This technique was widely applied to solve bounded systems 
\cite{Fait2,Heller}, and transmission phenomena through moving 
barriers\cite{Pimp}.

Such procedure does not avail with infinite barriers of potential, so to
avoid the infinite barrier at $q=0$, we have employed the symmetrized
potential 
\begin{eqnarray}
V(q,t)=\left\{ 
\begin{array}{ll}
0 & \mbox{if $0 < |q| \leq a(t)$ }\;, \\ 
V_0 & \mbox{if $ a(t) < |q|$ }\;.\label{pozo2}
\end{array}
\right. 
\end{eqnarray}
instead of potential (\ref{pozo1}).

We want to compute 
\begin{equation}
P_{\alpha}(t) = \int_{-a(t)}^{a(t)} |<\alpha,t_{0},t|q>|^{2} dq
\end{equation}
that is the probability to detect the particle inside the well at time $t$.

Because the system can be unbounded we will have a non stationary leaving
flux of probability going to $q=\pm \infty $. The spectral method
automatically imposes periodic boundary conditions, so the flux of
probability leaving at $q=+(-)q_{max}$ will appear as coming at 
$q=-(+)q_{max}$. So we need to kill the leaving flux. To get this we have
employed absorbing boundary conditions\cite{Kosloff} adding an absorptive
(purely imaginary) static potential barrier 
\begin{equation}
V_A=\frac{iU_0}{\cosh ^2{(\gamma (q-q_{max}))}}+\frac{iU_0}{\cosh ^2{(\gamma
(q+q_{max}))}}
\end{equation}
to the actual (real) time dependent potential well \ref{pozo2}. Here, $U_0$
and $\gamma $ are real parameters whose values were chosen to minimize the
reflection and transmission coefficients of $V_A$. All computations were
performed taking a time step $\Delta t=0.00064$ on a grid of 4096 steps of
spatial sampling $\Delta q=0.00125$. The absorbing region at the boundaries
includes 250 spatial steps. This fact determines the value of $\gamma $. We
have taken $U_{0}=14.79$. Figures ~\ref{well} summarizes the characteristic 
of the potential for the present calculation.

On the other hand we have change the sharp step function in the potential 
\ref{pozo2} by a soft version like a Saxon-Woods profile 
\begin{equation}
V(q,t)=\frac{V_0}{1+\exp {b(a(t)^2-q^2)}}
\end{equation}
as was prescribed in Ref.\ \onlinecite{Heller} to improve the convergence of
the numerical method. We have found that to ensure convergence and
stability, we need that the potential vary from $V_{0}/10$ to 
$9V_{0}/10$ in two spatial steps $\Delta q$. This determines the
parameter $b$.

\section{THE DECAY OF QUANTUM SYSTEMS}

\label{sec:decay}

To populate the quantum system in an equivalent way as in the classical
calculation, we consider the evolution of a quantum ensemble given by

\begin{equation}
\hat{\rho}(t_{0}) = \frac{1}{N} \sum_{\alpha \epsilon N}
|\alpha,t_{0},t_{0}><\alpha,t_{0},t_{0}|
\end{equation}
where $N$ is the number of eigenstates of the static well of deep 
$V_{0}=V(q,t=0)$ whose eigenenergies lie on the region that we have populated
the classical system. Such number depends on the value of $\hbar$. Setting
an appropriate value we calculate 
\begin{equation}
P(t) \equiv \int_{-a(t)}^{a(t)} <q| \hat{\rho}(t)|q> dq = \frac{1}{N}
\sum_{\alpha \epsilon N} P_{\alpha}(t) \;.  \label{proba}
\end{equation}
We have compute (\ref{proba}) taking $\hbar=0.0255$. For this value there
are ten bounded eigenstates of the static well whose eigenenergies are such
that $E_{min} \leq E \leq E_{max}$.

The results are shown in Figure ~\ref{qtdl} a) and b). The first shows a
log-lin plot where we can see that both curves, corresponding to harmonic
oscillations (curve A) and sawtooth oscillations (curve B) have
non-exponential behavior. The second one shows a log-log plot that evidences
an algebraic law from $t>25$ for both curves unlike to the classical
behaviors where for the same time interval they are qualitatively different
(see Fig.~\ref{cltdl}(a) and (b)). Let us remark that even algebraic 
($\propto 1/t^{\beta}$), the characteristic exponents of the qtdl are quite
different. The best fit for the qtdl corresponding to the harmonic
oscillation system is $\beta=1.45$ while it is $\beta=1.11 $ for the
sawtooth oscillation system.

\section{SUMMARY AND CONCLUSIONS}

\label{sec:con}

We have studied the time decay laws of two quasibounded systems in the
framework of classical and quantum mechanics. The classical results show
that the tdl's of the two systems are qualitatively different. One of the
tdl results exponential while the other is algebraic. This dissimilitude can
be explain in term of the diverse characteristics of the trapped periodic
orbits in the systems. On the other hand, the quantum systems do not show
such difference. We have obtained algebraic qtdl for both systems. However,
the characteristic exponents of the qtdl are different.

There are evidences that open quantum systems whose classical counterparts
are chaotic while they remain bounded (with tdl almost exponential) can
originate algebraic qtdl \cite{German1} with different characteristic
exponents. This effect can be explained by the distribution (Gaussian) of
resonances, assuming an exponential decay of each resonance and an initial
population of them. Varying the initial population, different characteristic
exponents can be obtained. Such algebraic decay does not seem to be related
to the existence of parabolic periodic orbits that generate algebraic tdl in
classical systems\cite{German}. In our systems, the initial population is
the same for both systems so we think that the different exponents can be
originate by the distribution of antiresonances (that is particular states
that are refractory to absorb energy from the oscillating wall) into each
well.

On the other hand, in Ref.\ \onlinecite{Lew,Har} it is shown that, in
general, the qtdl corresponding to chaotic systems have a power long time
tail ($\propto 1/t^{\beta}$). The specific characteristic exponent $\beta$,
depends on the number of open channels for the decay. In particular, for one
open channel the exponent is $\beta=1.5$\cite{Dit1}. This fact, could be
related to our result $\beta=1.45$ for the harmonic oscillating wall system.
Moreover, the characteristic exponent $\beta$, corresponding to non-chaotic
systems for one open channel and strong coupling between quasibounded states
and the continuum is closer to $\beta=1$\cite{Dit2} because the widening
of the distribution of resonance widths respect to those corresponding to 
chaotic systems. This 
result could be related to
our result $\beta=1.11$ in the case of sawtooth oscillating wall system.
These apparent connections are been studied.

To finish, in spite of the classical systems and the quantum systems
considered in the present work are not completely analogues (essentially
because the smoothing of the quantum potential), we expect that the tdl
corresponding to the classical systems with the smoothing potential will be
the same as the step potential systems. As we remark in Sec.\ \ref{sec:tdl},
the tdl for the classical systems is determined by the linear stability of
the trapped periodic orbits and such characteristic seems to be related to
the time dependence (instead the spatial dependence) of the potential.

\section*{ACKNOWLEDGMENTS}

This work was partially supported by UBACYT(Ex079). We would like to thank
E.Vergini for useful suggestions.

\appendix

\section{}

The present appendix is devoted to show the effect in the tdl when the very
low energy region is initially populated.

Let us consider a well of deep $V_{0}$. We have calculate the tdl starting
off the uniform population of whole trapped region (that is $0 \leq q \leq 1$
and $-\sqrt{2 V_{0}} \leq v \leq \sqrt{2 V_{0}}$. Fig.~\ref{alge}(a) shows
the tdl as a function of time. Here, we can see an algebraic behavior, that
is 
\begin{equation}
N(t)/N_{0} \propto 1/t^{\beta} \;,  \label{tdla}
\end{equation}
where $\beta=1$.

According Ref.\ \onlinecite{Majo1}, we expect a subset of periodic parabolic
orbits included in the bounded region. This subset corresponds to particles
that having $v\approx 0$ generate trajectories which are asymptotic to the
horizontal paths $v=0$ in the pipe version of our system.

To show this, let us consider $n(t)$ the fraction of initial conditions that
spend a time $t^{\prime}$ greater than a given time $t$, to reach the
oscillating wall. Looking at Fig. ~\ref{cano2}, we estimate, 
for $t>2/\sqrt{2 V_{0}}$, 
\begin{equation}
n(t) = \frac{1}{2 \sqrt{2 V_{0}}} \int dq \int_{-\tan{\ \phi_{min}}
=(1+q)/t}^{\tan{\phi_{max}}=(1-q)/t} d(tan {\phi})
\end{equation}
So that 
\begin{equation}
n(t) = \frac{1}{\sqrt{2 V_{0}}} \times \frac{1}{t}
\end{equation}
Following Ref.\ \onlinecite{Majo2}, we assume for the long time tail 
\begin{equation}
N(t)/N_{0} \approx \omega \int_{t^{\prime}=t}^{t^{\prime}= \infty}
g(t^{\prime}) dt^{\prime}
\end{equation}
where $g(t) dt$ is the fraction of initial conditions for the first
collision with the moving wall occurs between $t$ and $t+dt$ and $\omega$ is
the probability to leave the wall after one collision. As 
\begin{equation}
g(t) \approx \frac{-dn}{dt}
\end{equation}
we have the law \ref{tdla} for the tdl. We stress that in the case of 
Fig. ~\ref{tdla} the algebraic tail becomes for $t \approx 2/\sqrt{2 V_{0}}$ 
but in the general, if we populate the low energy region such that $|v| <
|v_{max}| < \sqrt{2 V_{0}}$ we expect that the tdl exhibits a crossover
between a stretched exponential and the algebraic decay law \ref{tdla} for
long times as it shows in Fig.~\ref{alge}(b).

When the initial population excludes the very low energy region, that is 
$|v_{min}| \leq |v| \leq |v_{max}|$, we have numerical evidence that the
algebraic tail for long times follows $1/t^{2}$ (See Fig. ~\ref{alge2}). So,
it differs by one respect to the precedent. This difference also occurs in
the decay of other quasibounded hamiltonian systems when a parabolic region
is initially populated or it is populated by an indirect way in \cite
{piko,Majo3}.

\newpage

\begin{figure}
\caption{a) One trajectory in the two dimensional pipe version of the
system. The coordinates are the time $t$ and the position of the particle 
$q$. b) The same trajectory in the billiard version. This point of view is
obtained by exploting the time lattice translation symmetry of the pipe
setting $\theta= t-[t/\tau] \tau$ as a coordinate.}
\label{cano}
\end{figure}

\begin{figure}
\caption{$E$ (kinetic energy) vs. $\Theta$ (reduced time) at $q=0$. 
Poincare
surface section corresponding to the system when the wall at $q=a(t)$
oscillates armonically. Both, $E$ and $\Theta$ are dimensionless and the
amplitude of the oscillation is $\delta=0.2$ (in units of $<a>_t$, 
see the text).}
\label{fermsec}
\end{figure}

\begin{figure}
\caption{One parabolic periodic orbit corresponding to the sawtooth
oscillation. Here $q$ is the position of the particle while $\theta$ is the
reduced time.}
\label{perio}
\end{figure}

\begin{figure}
\caption{$E$ (kinetic energy) vs. $\Theta$ at $q=0$. This Poincare surface
section shows only the low energy region and it was generated by one initial
condition. The solid straight lines limit the initial population that is
mentioned in the text. The dashed line indicates the deep of the well (it
separates the bounded region and the unbounded region). $E$ and $\Theta$ are
dimensionless.}
\label{fermsec2}
\end{figure}

\begin{figure}
\caption{a) Log-lin plot of the remaining population $N/N_{0}$ inside the
well against $t$ (tdl). We have fixed $\tau$ (the period of the moving wall)
as the unit of time and $<a>_{t}$ (the mean value of the position of the
moving wall) as the unit of length. (A) correspond to the harmonic
oscillations system and it has an almost exponential behavior while (B)
correspond to the sawtooth oscillations. In both cases the amplitude of the
oscillation is $\delta=0.2$ and the deep of the well is $V_{0}=8.63$. b) The
same curves of a) but in a log-log plot to show the algebraic behavior of
(B).}
\label{cltdl}
\end{figure}

\begin{figure}
\caption{This Figure shows the characteristic of the potential for the
computation of qtdl. The spatial grid has 4096 steps. The absorbing region
is marked by the shadow rectangles at the boundaries. Each side has 250
steps.}
\label{well}
\end{figure}

\begin{figure}
\caption{ a) Log-lin plot corresponding to the qtdl's $P$ vs. $t$. Curve (A)
corresponds to harmonic oscillation. Curve (B) corresponds to sawtooth
oscillations. b) Log-log plot for the same laws for $t>25$ to show the
algebraic behavior. The units are the same as Figure \protect\ref{cltdl}.}
\label{qtdl}
\end{figure}

\begin{figure}
\caption{Graph to show the fraction of initial conditions for which the
first collision on the moving wall occurs in times greater than $t$. Given
an initial position $q$, the velocities are $v=\tan{\phi}$ such $-\phi_{min}
\leq \phi \leq \phi_{max}$.}
\label{cano2}
\end{figure}

\begin{figure}
\caption{a) Log-log plot of the tdl corresponding to harmonic oscillating
well system starting off the uniform population of whole trapped region. The
deep of the well is $V_{0}= 0.78$ and $\delta=0.2$. We have also drawn the
straight line of slope one to clarify the characteristic exponent of the
algebraic behavior. b)Log-log plot of the tdl corresponding to harmonic
oscillating wall. In this case $V_{0}= 19.7$, $\delta=0.2$ and the initial
population is the same of a), that is $|v| \leq 1.25$. Here we can see the
algebraic behavior for the long time tail. We also draw the straight line of
slope one. The unit of time and the unit of length are the same as in 
Figure \protect\ref{cltdl}.}
\label{alge}
\end{figure}

\begin{figure}
\caption{ Log-log plot of the tdl corresponding to harmonic oscillating
wall. In this case $V_{0}=8.63$, $\delta=0.2$ and the initial population was
uniform in the region shown in the Figure \protect\ref{fermsec2} which 
excludes the
parabolic region. Here we can see the algebraic behavior for the long time
tail. We also draw the straight line of slope $2$. The unit of time and the
unit of length are the same as in Figure \protect\ref{cltdl}.}
\label{alge2}
\end{figure}
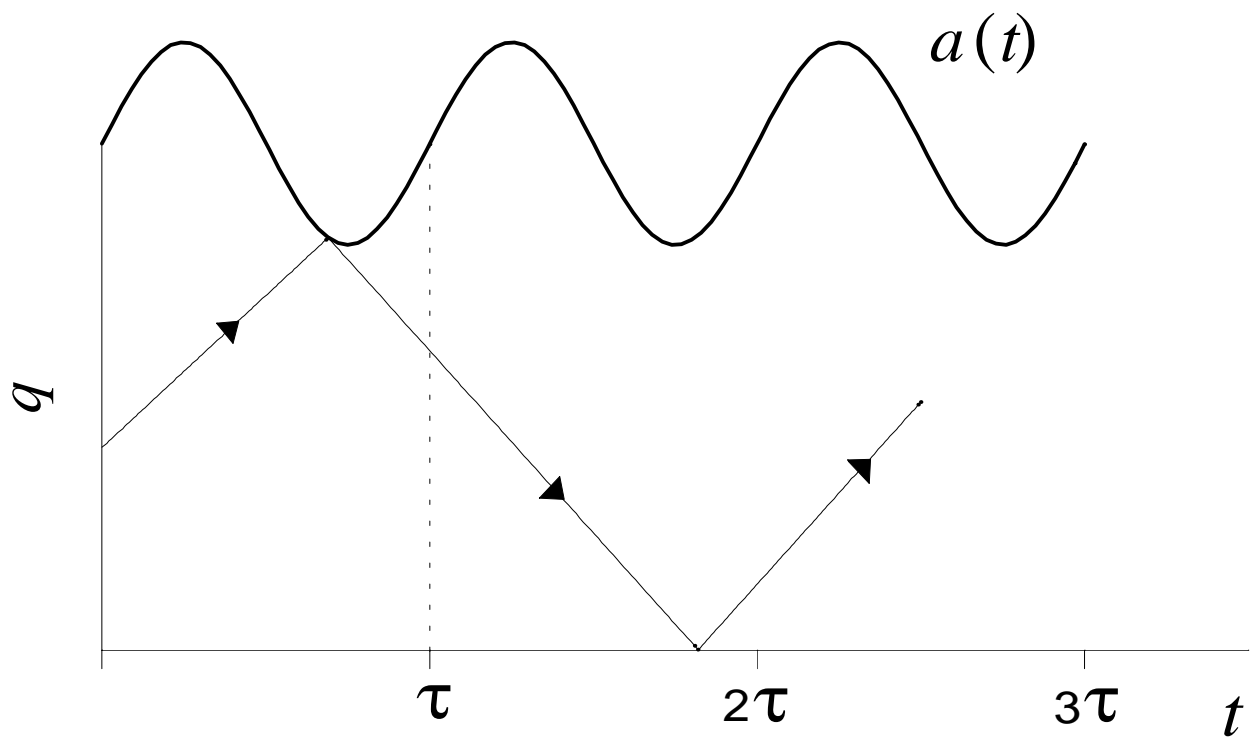
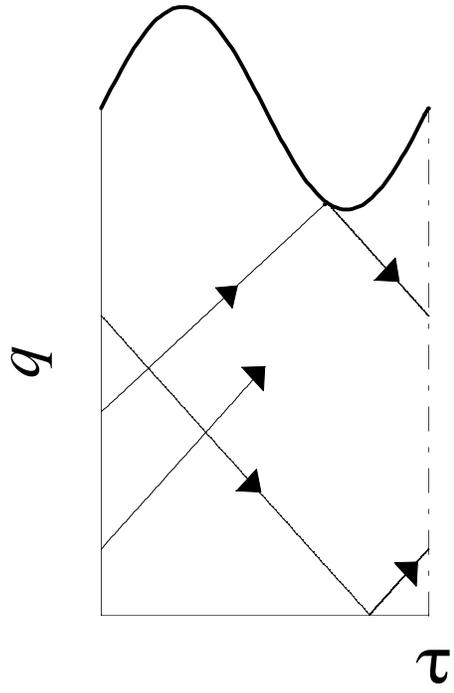
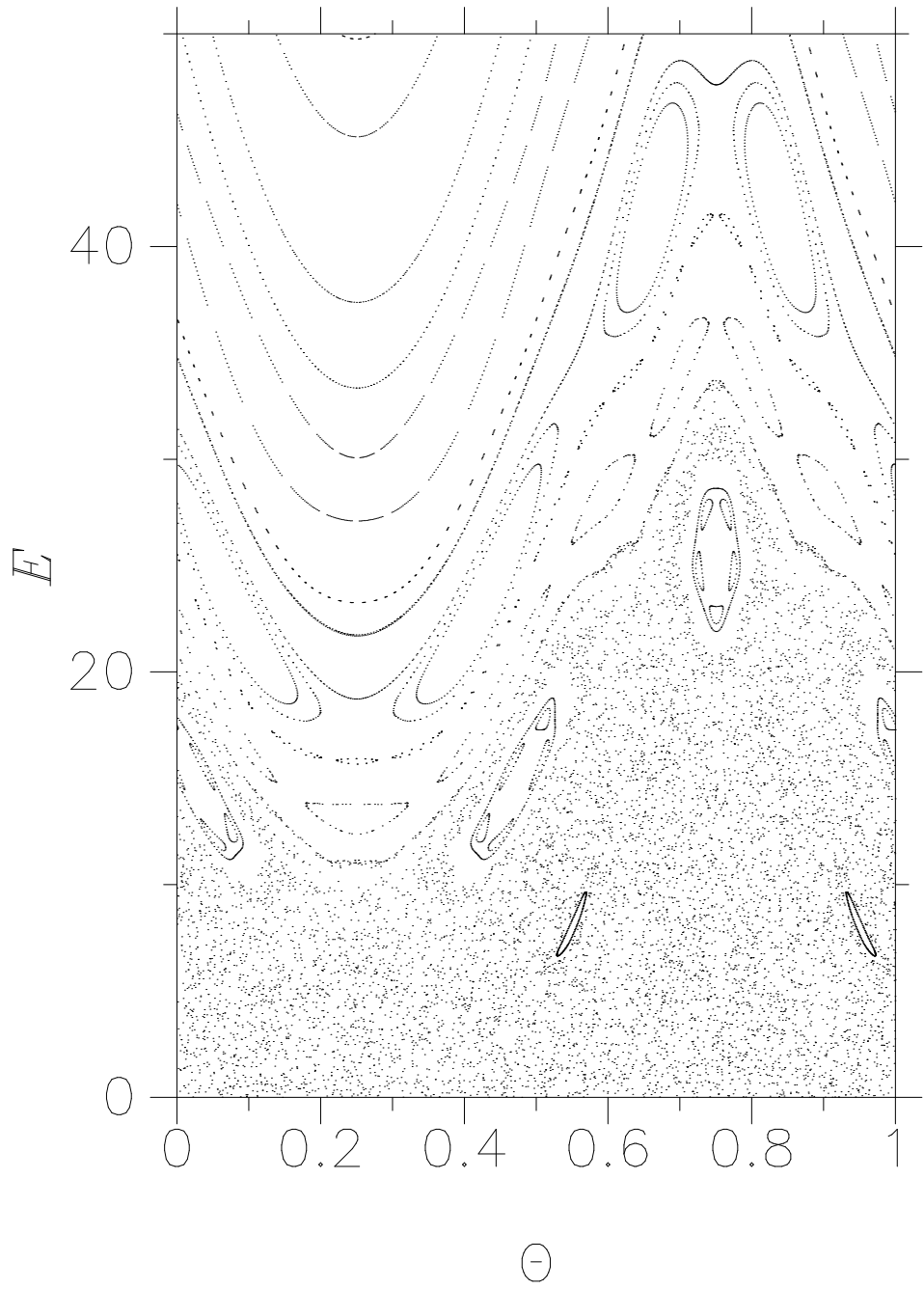
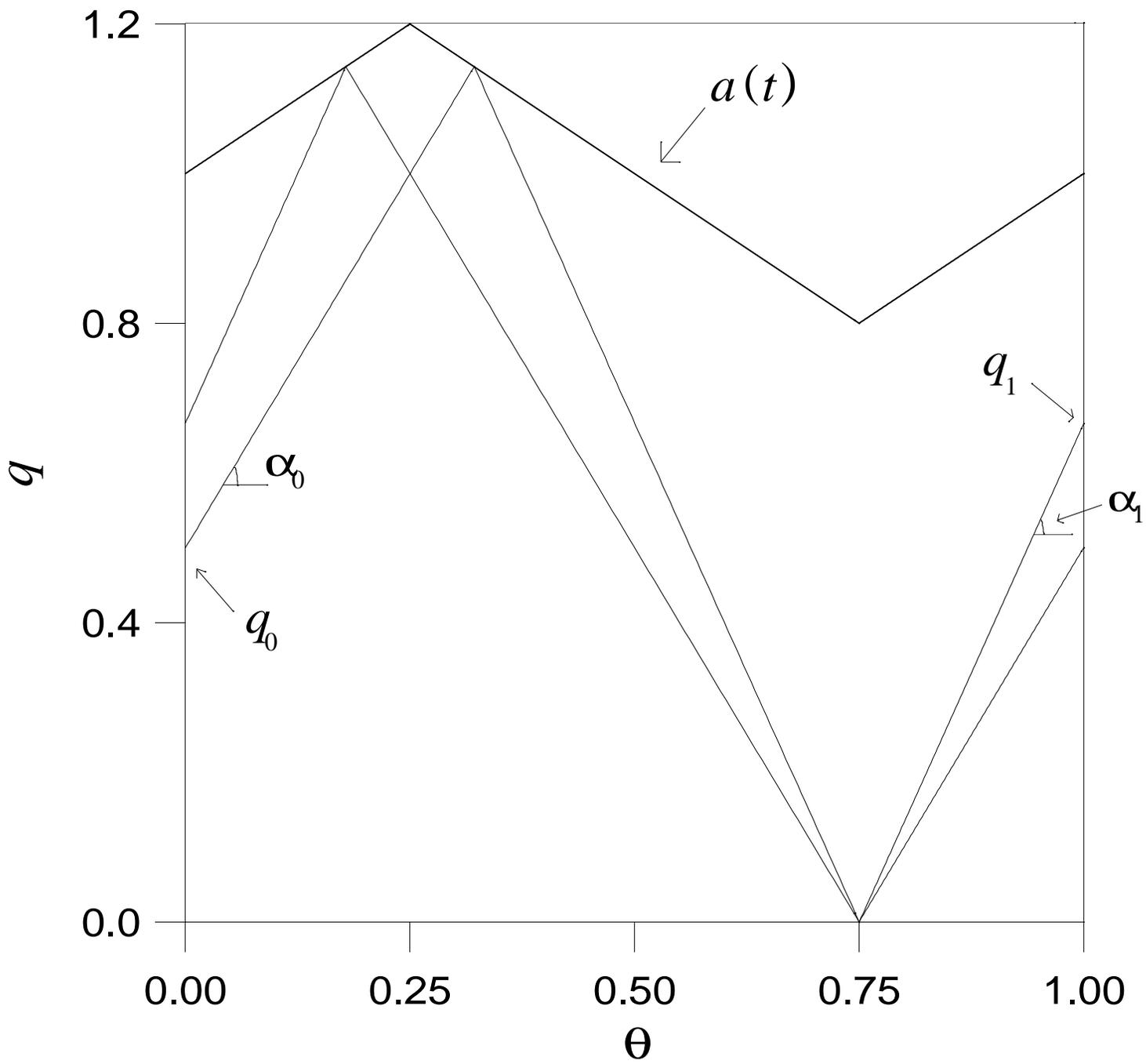
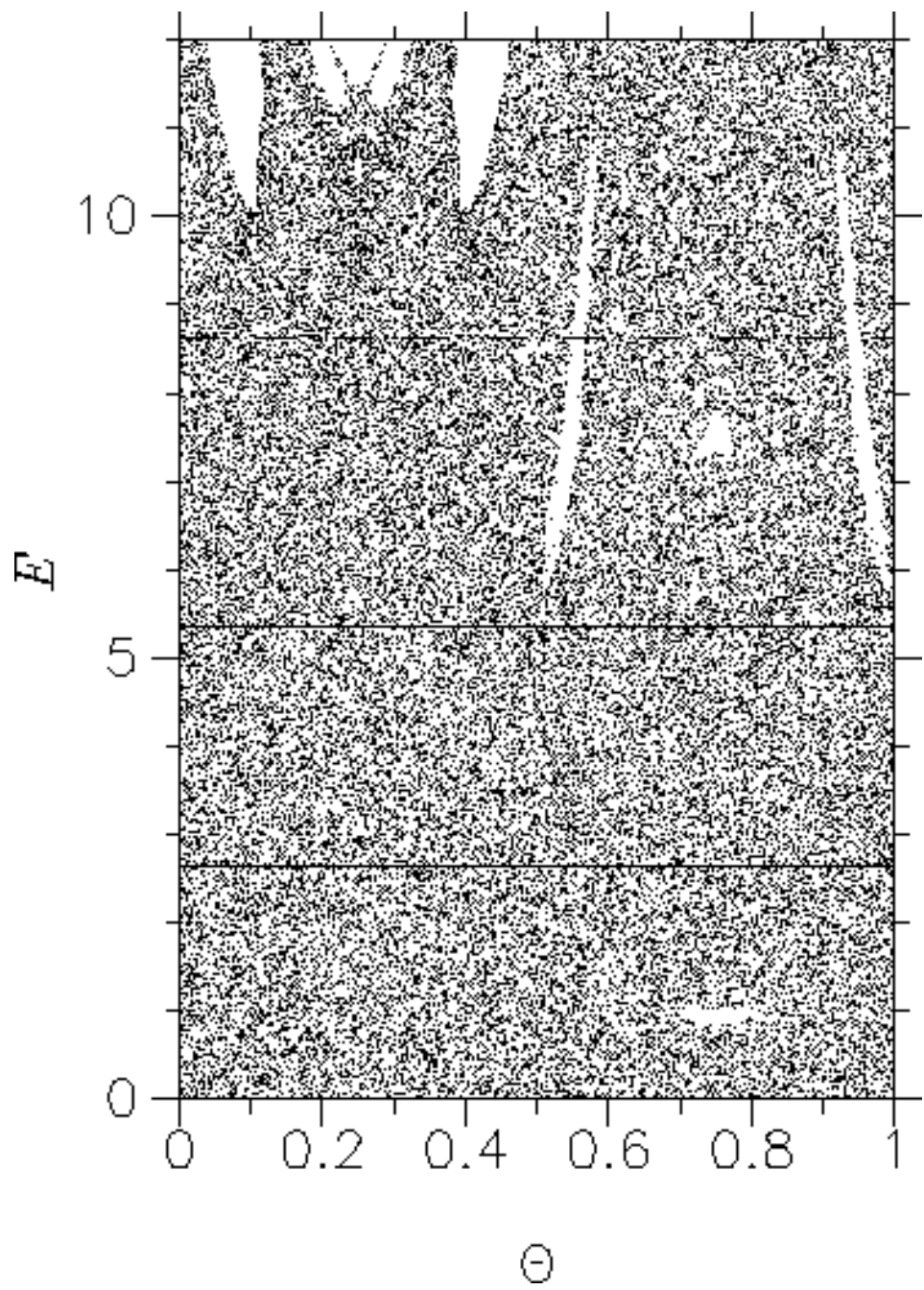
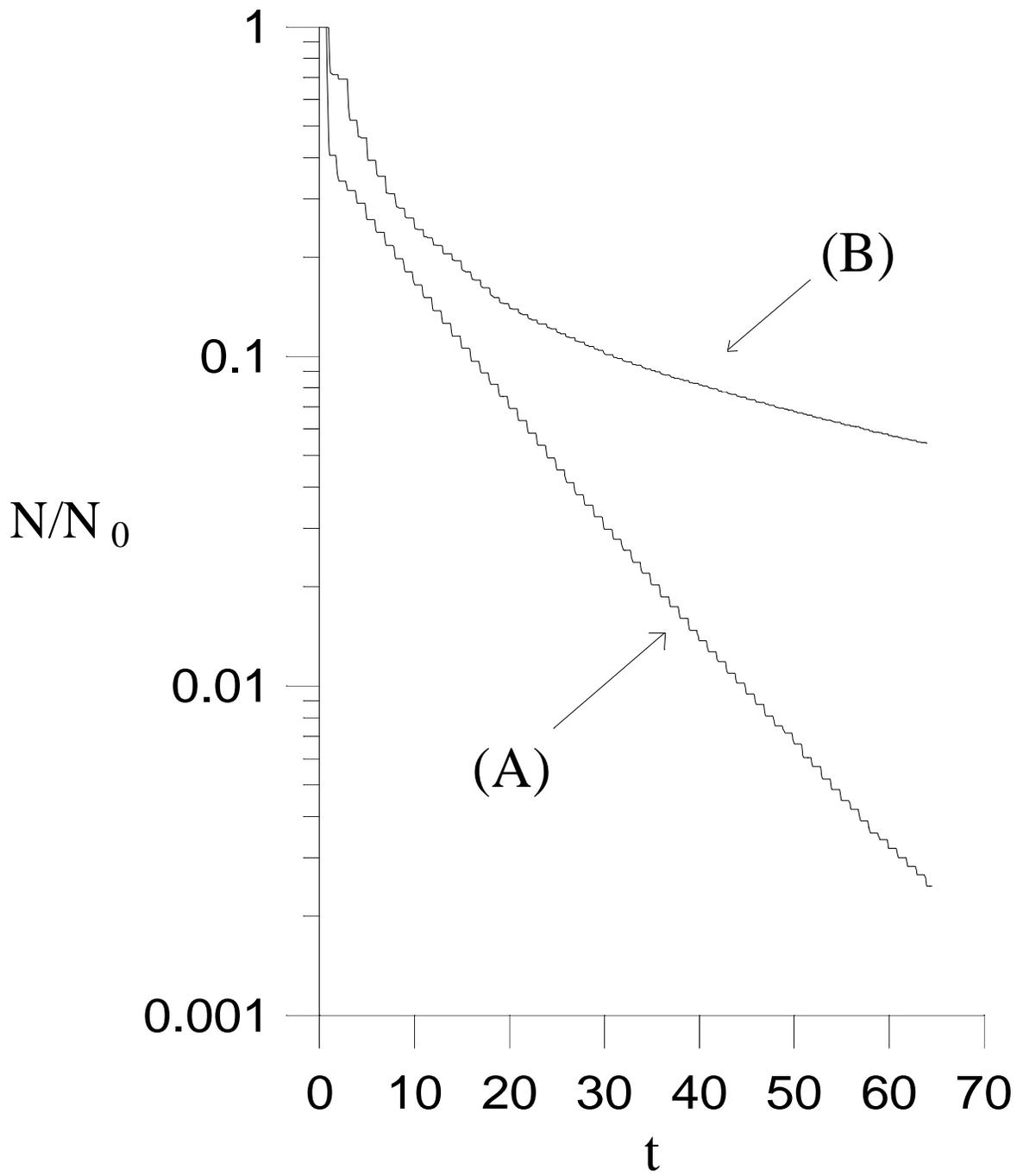
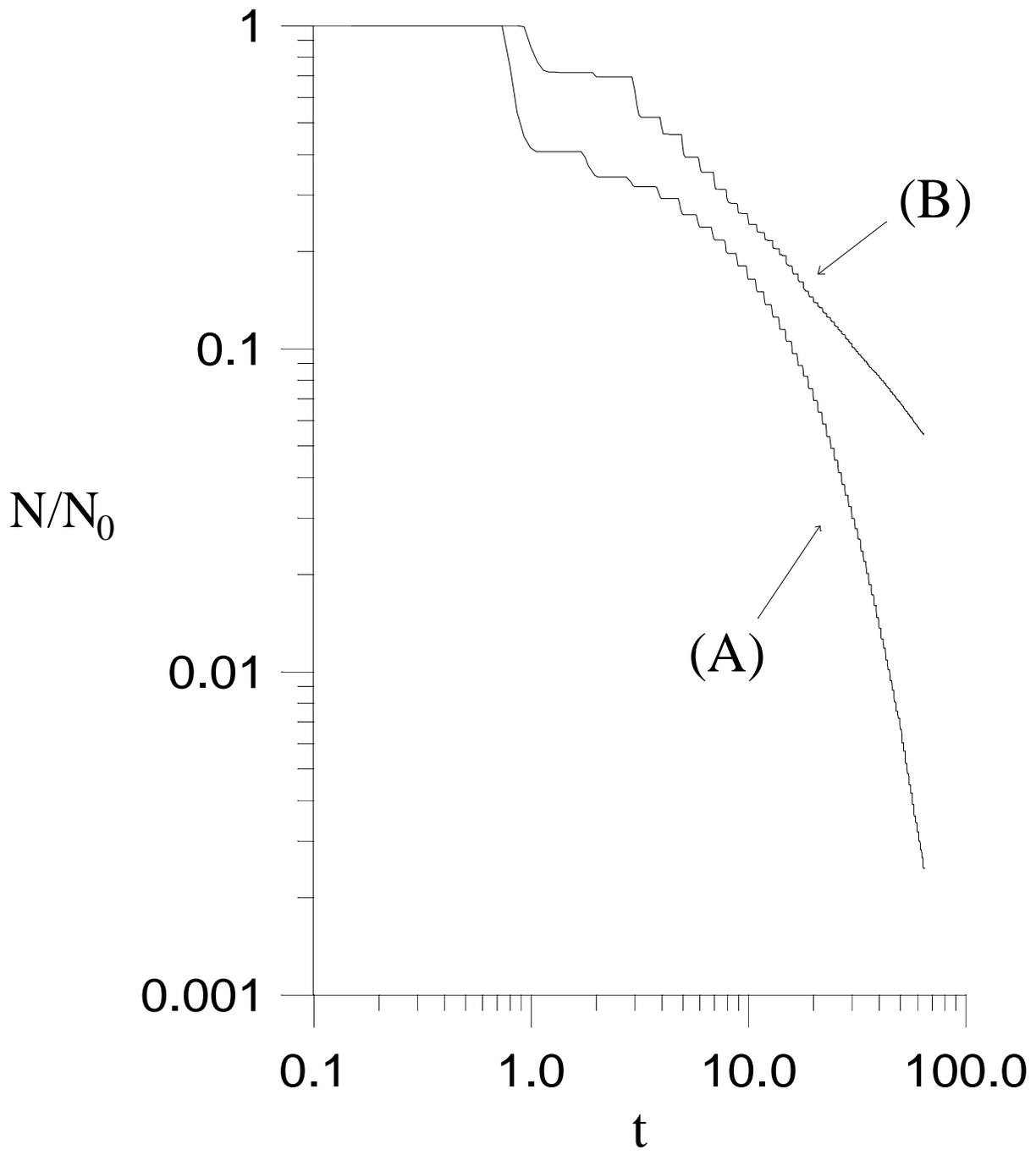
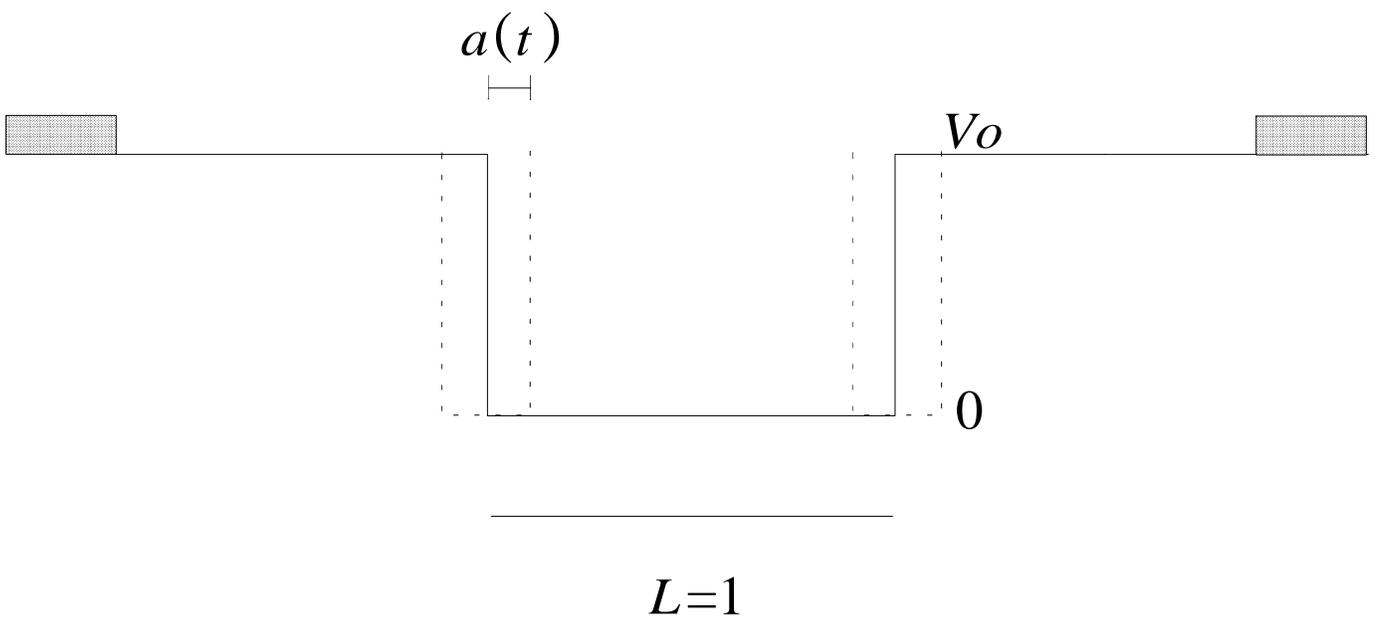
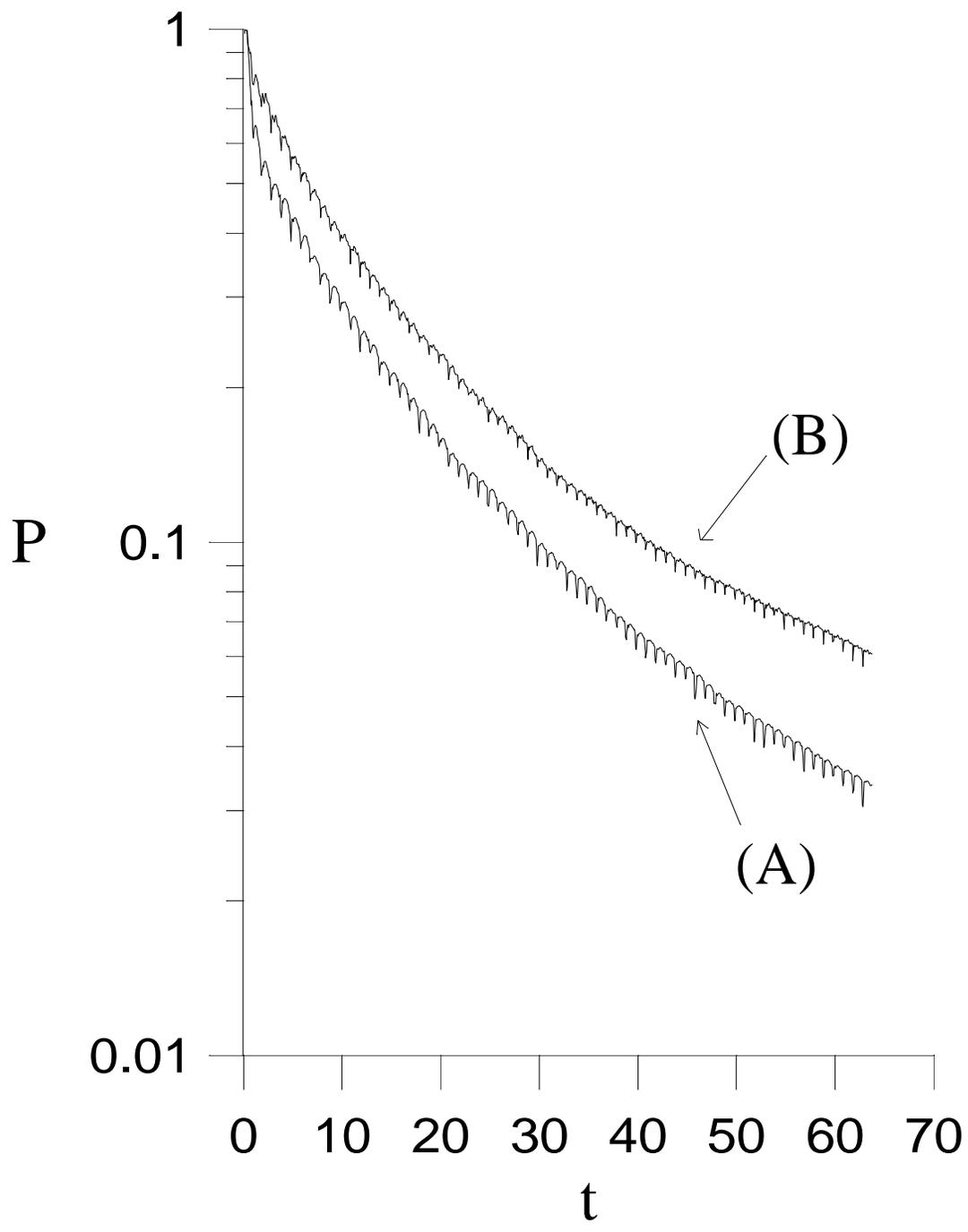
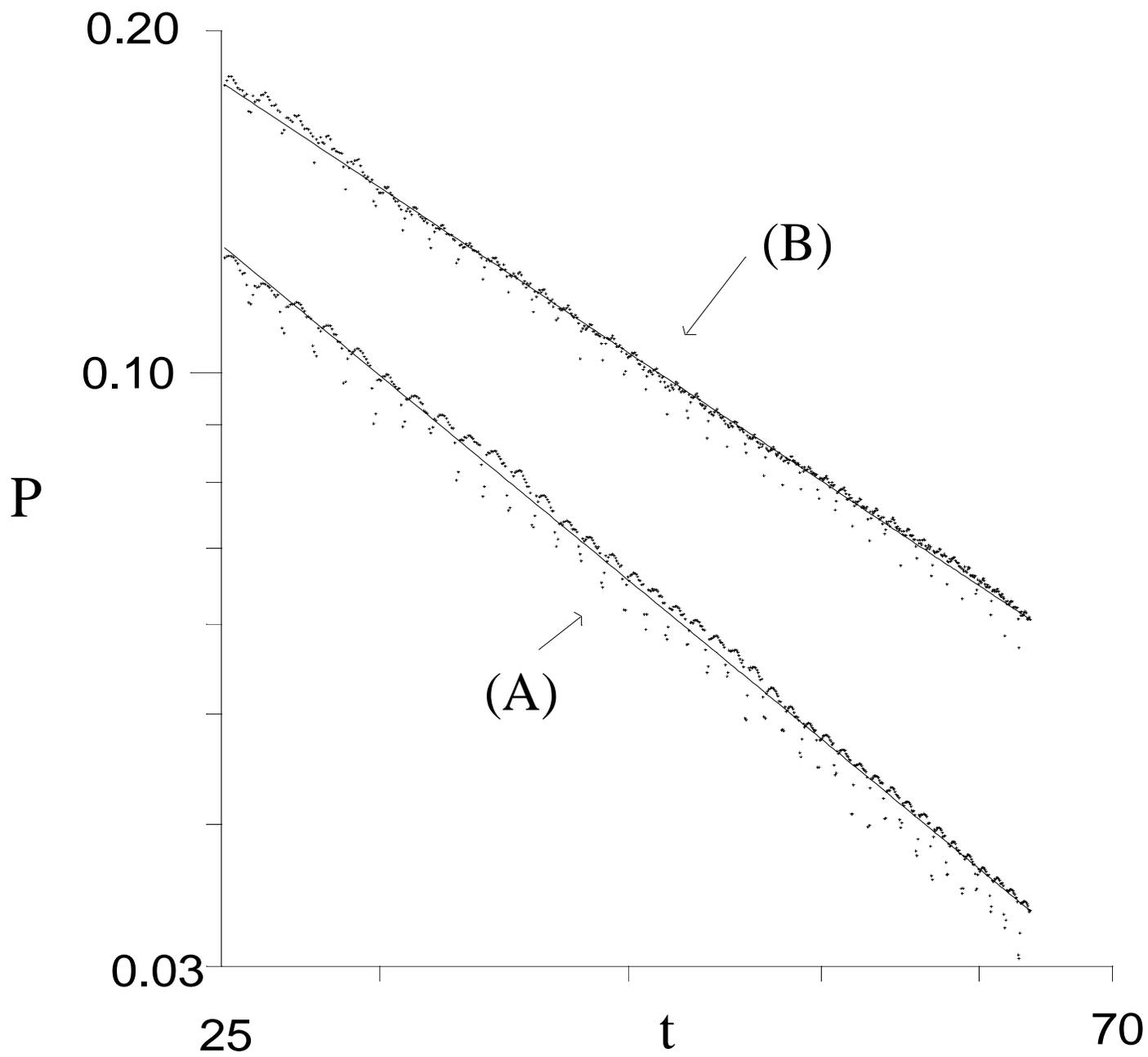
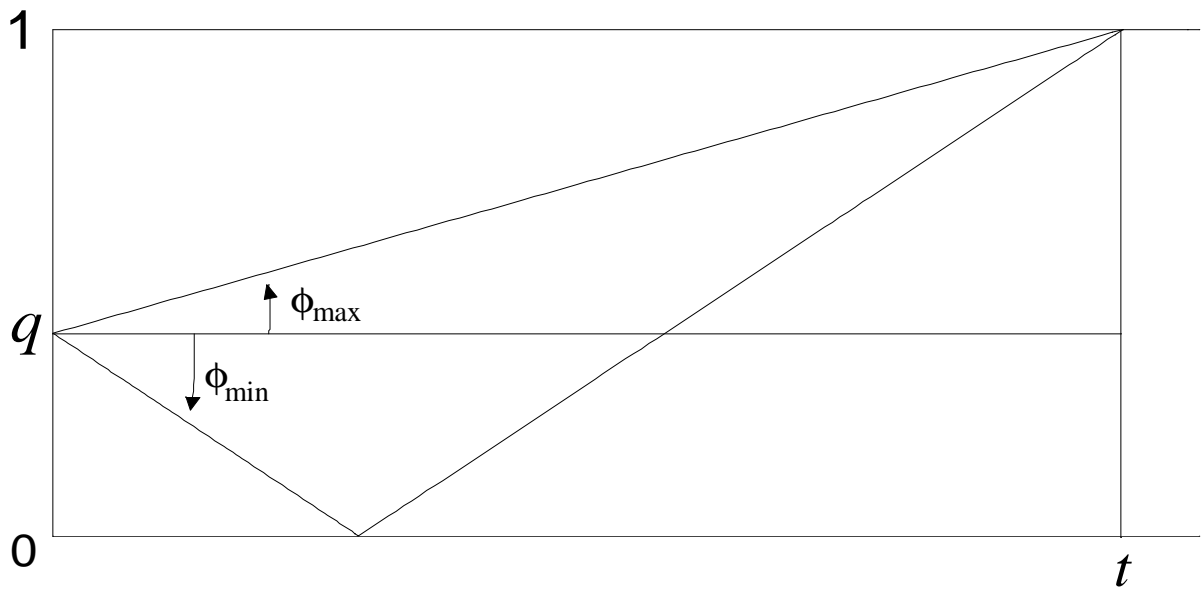
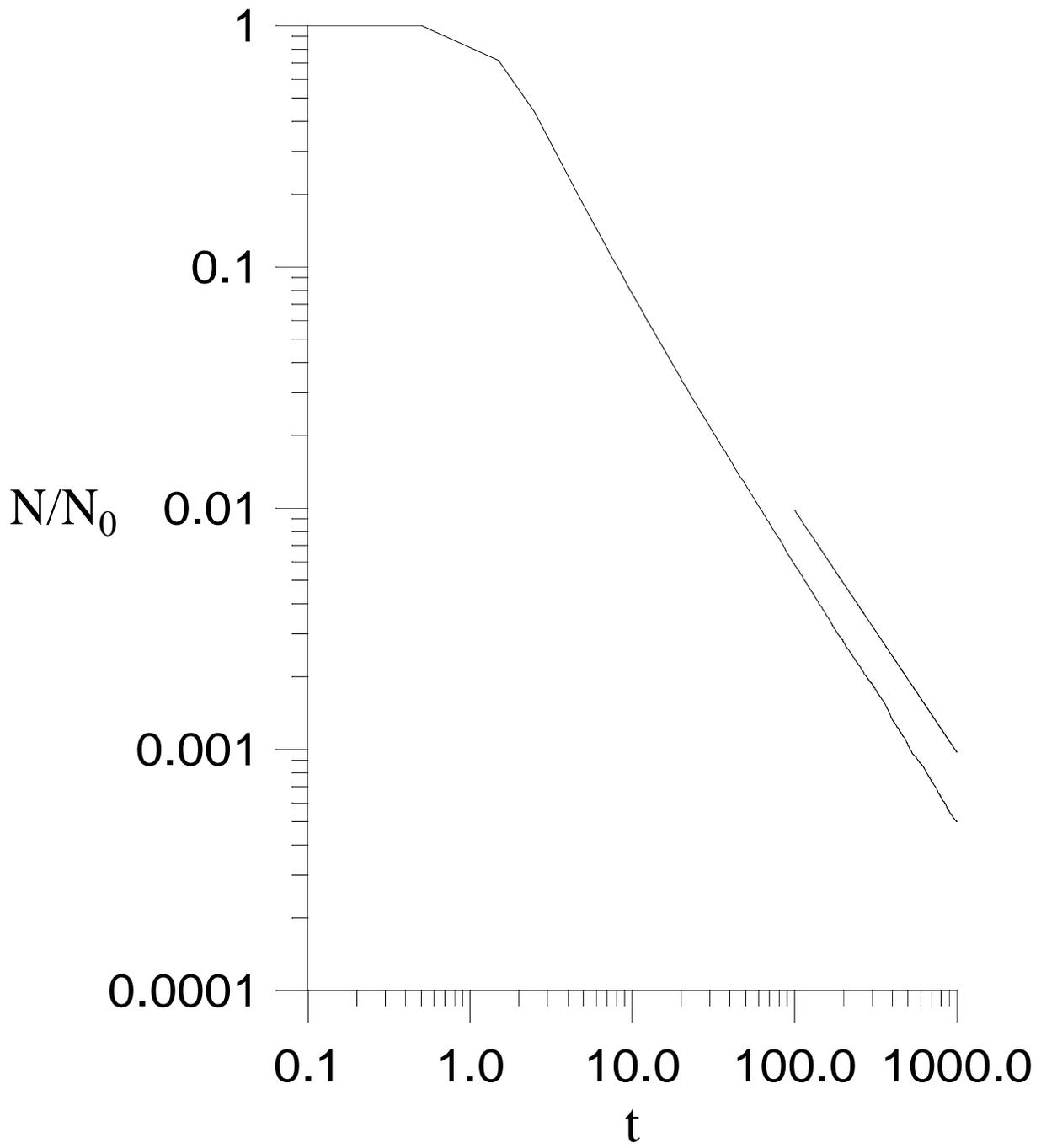
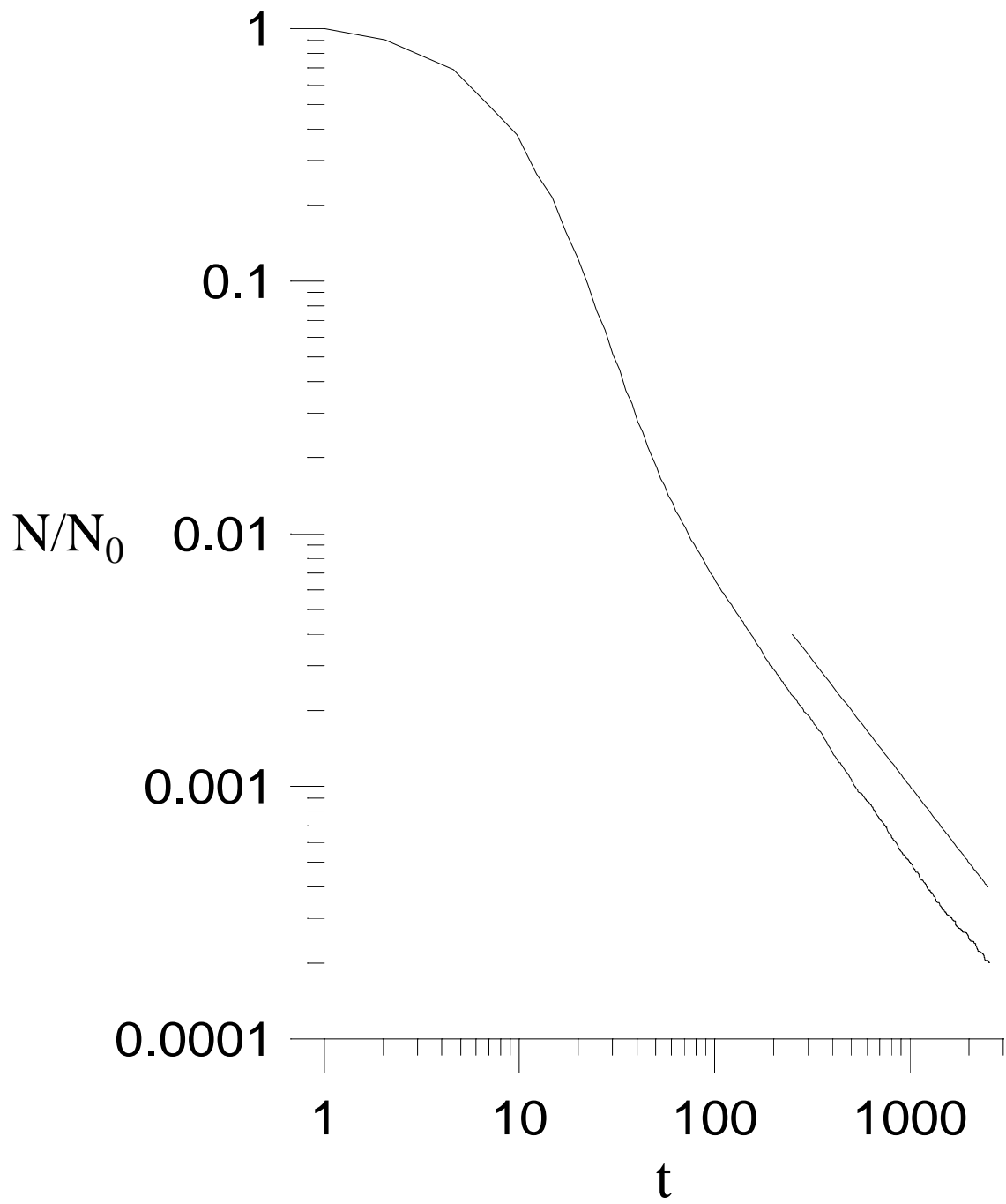
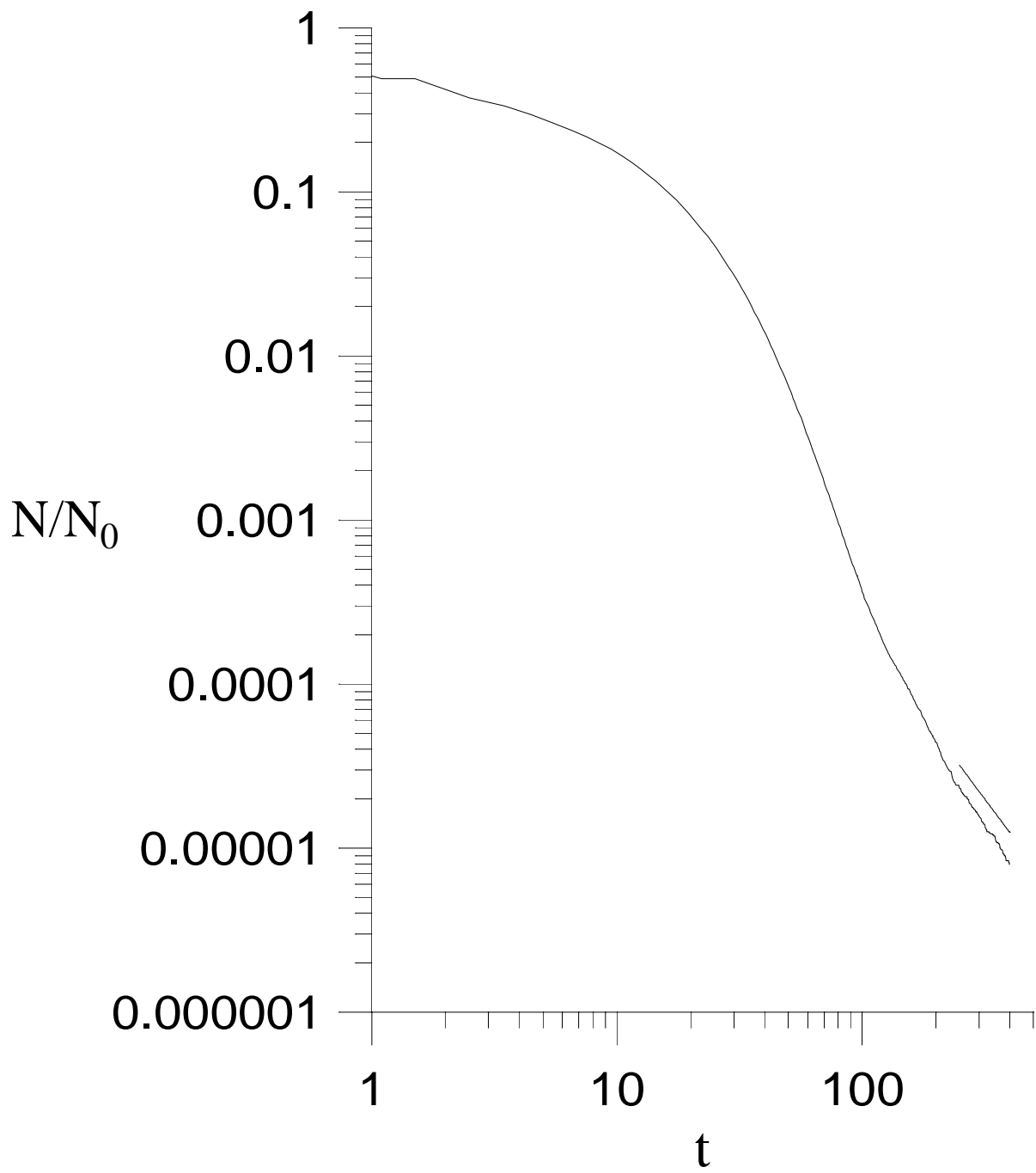
\end{document}